\documentclass[aps,amsfonts,nofootinbib]{revtex4-1}
\usepackage{epsfig}
\usepackage{graphicx}
\usepackage{slashed}
\usepackage{amsmath}
\usepackage{amsbsy}
\usepackage{bm}
\usepackage{color}
\usepackage{epsfig}
\newcommand{\ee}{\end{equation}}
\newcommand{\bb}{\begin{equation}}
\newcommand{\eqb}{\begin{eqnarray}}
\newcommand{\eqf}{\end{eqnarray}}

\def\sigmavec{\mbox{\boldmath$\sigma$}}

\newcommand{\1}{{\'{\i}}}
\def\sigmavec{\mbox{\boldmath$\sigma$}}

\def\1{\'{\i}}

\def\1{\'{\i}}
\begin{document}
\title{Analogies between nuclear physics and Dark Matter}
\author{D.  C\'arcamo}
\affiliation{Departmento de F\1sica, Universidad de Santiago de Chile, Casilla 307, Santiago, Chile}

\author{A. Riveros}
\affiliation{Departmento de F\1sica, Universidad de Santiago de Chile, Casilla 307, Santiago, Chile}

\author{J. Gamboa}
\affiliation{Departmento de F\1sica, Universidad de Santiago de Chile, Casilla 307, Santiago, Chile}

\begin{abstract}
A fermionic description of  dark matter using analogies with nuclear physics is developed.  At tree level, scalar and vector  processes are considered and the two-body potential are explicitly calculated using the Breit approximation.  We show that the total cross sections in both cases exhibit Sommerfeld enhancement.
\end{abstract}
\pacs{PACS numbers:}
\date{\today}
\maketitle
\section{Introduction}
Since the seventies, it is well known that in the center of the galaxies there is a high gamma radiation which can not be attributed nor to photons coming from cosmic rays neither to supernovas \cite{lista1}. The energy of photons has been observed and measured by different satellites and the line 511 cannot be explained by using a  conventional point of view \cite{Johnson}.

Nevertheless there are also other interesting results in particle physics related to the problem mentioned above which require further explanation. In this direction, it seems that the line of 511 keV would find a natural explanation in the last years (although not definitive considering some others explanations, but certainly the most convincing) \cite{Finkbeiner_Weiner}. Basically the great production of electron-positron pairs can be attributed to the dark matter annihilation \cite{Strong_etal}-\cite{Dobler_Finkbeiner}  through processes that are beyond the standard model \cite{Finkbeiner_Weiner}-\cite{Cholis_etal}.

A possible explanation is to argue that the total cross section of annihilation would be enlarged in such a way to be consistent with observations. This mechanism has been proposed by Arkani-Hamed et. al.  \cite{arkani,iengo1,iengo2} and it is direct consequence of the so called Sommerfeld enhancement (${\cal S}$), i.e. the assumption that 
the ratio between the probability densities of the wave function in ${\bf x}=0$ e ${\bf x}=\infty$ are normalized to the ${\cal S}$ factor \cite{Sommerfeld}.

As the dark matter is essentially non-baryonic and interacts only at low energies, one can establish analogies with nuclear physics which can then be extrapolated and be valid at least as effective descriptions \cite{Bertulani}. 

In this context neutrinos seen as cold dark matter \cite{paola} are the non-relativistic analogue of nucleons and therefore,  the phenomenological potentials of nuclear physics should be  also useful in the dynamic description of the cold dark matter neutrinos. 
In this paper we will explore this possibility and we we will show how these ideas fit with some of the results discussed in dark matter physics.

The paper is organized as follows: in section II we discuss some basics issues about the phenomenological potential in nuclear physics and dark matter and  the notation is established, in section III include the calculation of the phenomenological potential in terms of the Breit procedure, in section IV we give some examples where 
the Sommerfeld enhancement is calculated and  finally, section V contains the conclusions.

\section{Nuclear Potential and Dark matter }
In the context of the early nuclear physics, the nucleons interact by interchanging mesons \cite{Bertulani}. But herein nuclear physics itself, is an effective description of a fundamental theory, namely QCD.

In this effective description the non-relativistic nucleon-nucleon potential is phenomenological and it has the general form 

\begin{equation}
V=V_{s}(r)  +  \sum_{i} V_i(r) O_{i}, \label{1}
\end{equation}
where $V_{s}(r)$ is the part of the potential wich only depends on the relative distance $r$ between the interacting nucleons, $O_{i}$ is the operators for spin contribution and therefore $V_i(r) O_{i}$ are the terms of the potential which are obtained in the low energy expansion. 
We have in the coordinates space: 

\begin{eqnarray*}
 O_{1}&=&{\bf \sigmavec_{1}} \cdot {\bf \sigmavec_{2}},  \\
O_{2}&=&({\bf \sigmavec_{1}} \cdot {\bf x})( {\bf \sigmavec_{2}} \cdot {\bf x}),  \\
O_{3}&=& {\bf L}   \cdot {\bf S}, \\
O_{4}&=& ({\bf L}   \cdot {\bf S})^n,
\end{eqnarray*}
where ${\bf S}=\frac{1}{2}({\bf \sigmavec_{1}}+{\bf \sigmavec_{2}})$ is the total spin and $L$ is the orbital angular momentum (we will use $\hbar =1$ in the follow).

These operators $O_{i}$ are known as  spin-spin, tensor, spin-orbit and  higher order of spin-orbit respectively and which can be written in terms of the total spin operator  ${\bf S}$.

The spin-spin term is ${\bf \sigmavec_{1} \cdot \sigmavec_{2}}=2 {\bf S}^2-3$; therefore the spin-orbit term is ${\bf L\cdot S}=({\bf J}^2 -{\bf L}^2 -{\bf S}^2)/2$, where ${\bf J}={\bf L}+{\bf S}$ is the total angular momentum. The tensor interaction is given by  $({\bf \sigmavec_{1} \cdot x})({\bf \sigmavec_{2} \cdot x})=2 ({\bf S \cdot x})^2-r^2$.

The explicit form of (\ref{1}) is not derived directly because we do not know exactly the nature of the effective interaction (without invoking QCD), therefore only considering the effective character of this description and  using some reasonable assumptions, we can find explicit forms for (\ref{1}). 

Although this strategy is commonly used in nuclear physics, is particularly interesting to explore in detail to study new effects in the dynamics of dark matter. This will be the purpose of the next section.



 \section{The Breit equation, massive bosons and Dark Matter} 
 The processes involving dark matter are  low energy ones and it can --at least in first approximation-- to exchange scalar or vector bosons and if, as our description is effective, also can be considered as massive bosons. A potentially dangerous point for massive fields a quantum field theory is the lack of renormalizability and unitarity but at this effective level this is 
 circumvented here because we will consider only very low energy processes and at the tree level. 
\vskip 0.5cm
Let us start considering the scattering amplitude for two particles interacting by a massive vector boson 
\begin{equation}
M_{fi}=\bar{u}'_{1}\gamma^{\mu} u_{1} D_{\mu \nu}(q)\bar{u}'_{2}\gamma^{\nu} u_{2}, \hspace{5mm} q=p'_{1}-p_{1}=p_{2}-p'_{2},
\end{equation}
where $\gamma^{\mu}$ are the gamma matrix. 

The propagator for a vector boson with mass $Q_c$ ($Q_c=Mc$)is given by
\begin{equation}
D_{\mu \nu}(q)=\frac{1}{q^2-Q_{c}^2} \left( -\eta_{\mu \nu} +\frac{q_{\mu}q_{\nu}}{Q_{c}^2}  \right).
\end{equation}
\begin{figure*}[ht]
\centering
\includegraphics[width=0.2\textwidth]{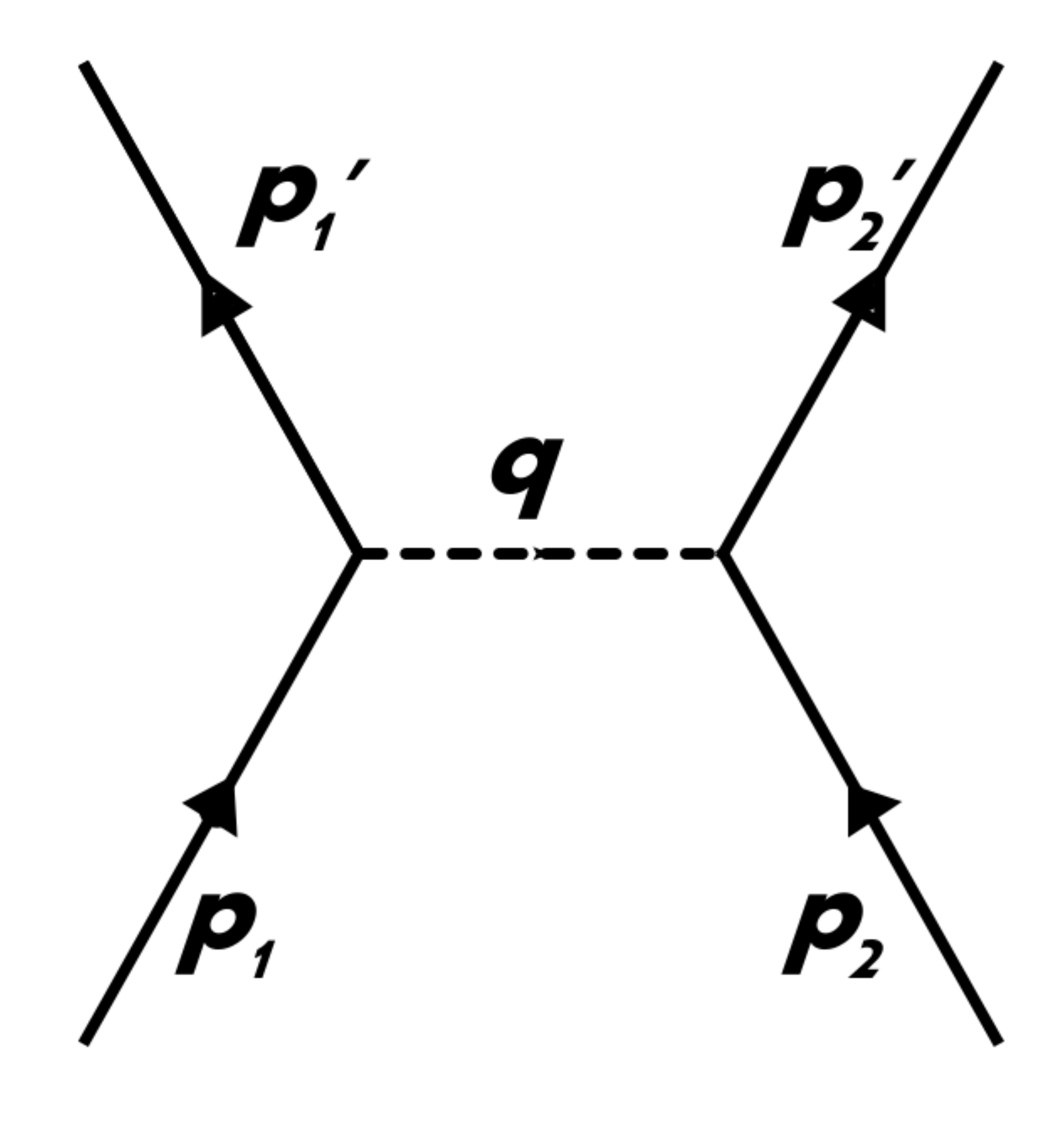}
\caption{fermion-fermion scattering}
\end{figure*}

The last expression can be written in components as follows
\begin{equation}
D_{00}(q)=\frac{1}{(\omega/c)^2-{\bf q}^2-Q_{c}^2} \left( -1 +\frac{(\omega/c)^2}{Q_{c}^2}  \right), \hspace{3mm} D_{0i}(q)=\frac{1}{(\omega/c)^2-{\bf q}^2-Q_{c}^2} \left(\frac{\omega q_{i}/c}{Q_{c}^2}  \right), \nonumber
\end{equation}
\begin{equation}
 D_{ik}(q)=\frac{1}{(\omega/c)^2-{\bf q}^2-Q_{c}^2} \left(\delta_{ik} +\frac{q_{i}q_{k}}{Q_{c}^2}  \right),
\end{equation}
where we use the convention $\eta_{00}=$1.

In order to obtain the low energy limit of the Dirac equation, we decompose  the bispinors into spinors components and until order  $1/c^2$ approximation, then \cite{bere}:

\begin{equation}
u({\bf p})=\sqrt{2m}
\begin{bmatrix}
 \left(1-\frac{{\bf p}^2}{8m^2c^2}  \right) w \\
\frac{{\bf \sigmavec} \cdot {\bf p}}{2mc} w
\end{bmatrix},\hspace{5mm} \bar{u}=u^{\ast}\gamma^{0},
\end{equation}
where $w$ is the amplitude of the Schr\"odinger  plane wave.

Thus, in this approximation, the terms ${\bar u} \gamma^\mu u$ in temporal and spatial components are 

\eqb 
\bar{u}'_{1}\gamma^{0} u_{1}&=& 2m_{1}w'^{\ast}_{1} w_{1}\left(1-\frac{{\bf q}^2}{8m^{2}_{1}c^2}+\frac{i{\bf \sigmavec_{1}} \cdot [{\bf q} \times {\bf p_{1}}]}{4m^{2}_{1}c^2}       \right), \nonumber 
\\
\bar{u}'_{1}\gamma^{i} u_{1}&=&\frac{w'^{\ast}_{1} w_{1}}{c} \left(2p^{i}_{1}+q^{i}+i[ {\bf \sigmavec_{1}} \times {\bf q}]^{i}  \right). \nonumber
\eqf

Solving $u_{2}$ is similar but doing the change ${\bf q} \rightarrow -{\bf q}$, because ${\bf q}={\bf p'_{1}}-{\bf p_{1}}={\bf p_{2}}-{\bf p_{2}'}$.
\eqb
\bar{u}'_{2}\gamma^{0} u_{2}&=&2m_{2}w'^{\ast}_{2} w_{2}\left(1-\frac{{\bf q}^2}{8m^{2}_{2}c^2}-\frac{i{\bf \sigmavec_{2}} \cdot [{\bf q} \times {\bf p_{2}}]}{4m^{2}_{2}c^2}       \right), \nonumber
\\
\bar{u}'_{2}\gamma^{i} u_{2}&=&\frac{w'^{\ast}_{2} w_{2}}{c} \left(2p^{i}_{2}-q^{i}-i[ {\bf \sigmavec_{2}} \times {\bf q}]^{i}  \right). \nonumber
\eqf

Furthermore, in this approximation, the propagators can be written as: 

\begin{equation}
D_{00}(q)=\frac{1}{{\bf q}^2+Q_{c}^2} , \hspace{3mm} D_{0i}(q)=0, \hspace{3mm}  D_{ik}(q)=\frac{-1}{{\bf q}^2+Q_{c}^2} \left(\delta_{ik} +\frac{q_{i}q_{k}}{Q_{c}^2}  \right).
\end{equation}

In order to clarify the calculation, we separate each term and then we take into account only terms order until $1/c^2$.


\begin{eqnarray*}
 \bar{u}'_{1}\gamma^{0} u_{1} D_{0 0}(q)\bar{u}'_{2}\gamma^{0} u_{2} 
   &=&2m_{1}2m_{2}w'^{\ast}_{1} w_{1} w'^{\ast}_{2} w_{2} \Biggl[  \frac{1}{{\bf q}^2+Q_{c}^2} -\frac{1}{8c^2}\left(  \frac{1}{m^2_{1}}+ \frac{1}{m^2_{2}} \right)  \frac{{\bf q}^2}{{\bf q}^2+Q_{c}^2}+\\
   && {}+\frac{i}{(2m_{1}c)^2} \frac{{\bf \sigmavec_{1}} \cdot [{\bf q} \times {\bf p_{1}}]}{{\bf q}^2+Q_{c}^2}-\frac{i}{(2m_{2}c)^2}\frac{{\bf \sigmavec_{2}} \cdot [{\bf q} \times {\bf p_{2}}]}{{\bf q}^2+Q_{c}^2}   \Biggr] , 
\end{eqnarray*}


\begin{eqnarray*}
 \lefteqn{\bar{u}'_{1}\gamma^{i} u_{1} D_{i k}(q)\bar{u}'_{2}\gamma^{k} u_{2}= }\\
  &=& -\frac{w'^{\ast}_{1} w_{1} w'^{\ast}_{2} w_{2}  }{c^2} \Biggl[ \frac{4{\bf p_{1}} \cdot  {\bf p_{2}}}{{\bf q}^2+Q_{c}^2}+\frac{2({\bf p_{2}}-{\bf p_{1}}) \cdot  {\bf q}-{\bf q}^2}{Q_{c}^2}+\frac{2i{\bf \sigmavec_{1}} \cdot [  {\bf q} \times {\bf p_{2}}]}{{\bf q}^2+Q_{c}^2}-\frac{2i{\bf \sigmavec_{2}} \cdot [ {\bf q}\times {\bf p_{1}}]}{{\bf q}^2+Q_{c}^2}+\\
   && {} + {\bf \sigmavec_{1}} \cdot {\bf \sigmavec_{2}} \frac{{\bf q}^2}{{\bf q}^2+Q_{c}^2}-\frac{({\bf \sigmavec_{1}} \cdot {\bf q})({\bf \sigmavec_{2}} \cdot {\bf q})}{{\bf q}^2+Q_{c}^2}) \Biggr],
\end{eqnarray*}
 where the following identities have been used 
\begin{equation}
({\bf \sigmavec_{1}} \times {\bf q}) \cdot  ({\bf \sigmavec_{2}} \times {\bf q})=({\bf \sigmavec_{1}} \cdot {\bf \sigmavec_{2}}){\bf q}^2 - ({\bf \sigmavec_{1}} \cdot {\bf q})({\bf \sigmavec_{2}} \cdot {\bf q}), \hspace{3mm} ({\bf a} \times {\bf b}) \cdot {\bf c} = {\bf a} \cdot ({\bf b} \times {\bf c}).
\end{equation}

Replacing these terms in the general expression of scattering amplitude, we obtain 

\begin{equation}
M_{fi}=2m_{1}2m_{2} w'^{\ast}_{1}w'^{\ast}_{2}  U({\bf p_{1}},{\bf p_{2}},{\bf q})    w_{1}w_{2},
\end{equation}
where $U({\bf p_{1}},{\bf p_{2}},{\bf q})$ is the two-body Breit potential, {\it i.e.} 

\begin{eqnarray}
 \lefteqn{U({\bf p_{1}},{\bf p_{2}},{\bf q})= } \nonumber\\
   &=&    \Biggl[ \frac{1}{{\bf q}^2+Q_{c}^2} -\frac{1}{8c^2}\left(  \frac{1}{m^2_{1}}+ \frac{1}{m^2_{2}} \right)  \frac{{\bf q}^2}{{\bf q}^2+Q_{c}^2} -\frac{1}{m_{1}m_{2}c^2}\frac{{\bf p_{1}} \cdot  {\bf p_{2}}}{{\bf q}^2+Q_{c}^2} +\frac{1}{2m_{1}m_{2}c^2}\frac{({\bf p_{2}}-{\bf p_{1}}) \cdot  {\bf q}-{\bf q}^2}{Q_{c}^2} \nonumber\\  
   && {}+\frac{i}{(2m_{1}c)^2}\frac{{\bf \sigmavec_{1}} \cdot [{\bf q} \times {\bf p_{1}}]}{{\bf q}^2+Q_{c}^2}-\frac{i}{(2m_{2}c)^2} \frac{{\bf \sigmavec_{2}} \cdot [{\bf q} \times {\bf p_{2}}]}{{\bf q}^2+Q_{c}^2}-\frac{i}{2m_{1}m_{2}c^2}\frac{ {\bf \sigmavec_{1}} \cdot [ {\bf q} \times {\bf p_{2}}]}{{\bf q}^2+Q_{c}^2}+ \nonumber\\
   && {}+\frac{i}{2m_{1}m_{2}c^2}\frac{{\bf \sigmavec_{2}}  \cdot [ {\bf q} \times {\bf p_{1}}]}{{\bf q}^2+Q_{c}^2}-\frac{1}{4m_{1}m_{2}c^2} {\bf \sigmavec_{1}} \cdot {\bf \sigmavec_{2}} \frac{{\bf q}^2}{{\bf q}^2+Q_{c}^2}+  \frac{1}{4m_{1}m_{2}c^2}\frac{({\bf \sigmavec_{1}} \cdot {\bf q})({\bf \sigmavec_{2}} \cdot {\bf q})}{{\bf q}^2+Q_{c}^2} \Biggr]. \nonumber\\
\end{eqnarray}

which shows that the  lowest order interaction is given by the Yukawa  potential.


We see that if $m_{i} \ll M$ (or the same  $ |{\bf q}| \ll Q_{c}$, light dark matter), then we will have to first order $\frac{1}{{\bf q}^2+Q_{c}^2}\simeq \frac{1}{Q_{c}^2}$. 

In order to obtain the potential in the coordinate-space we need to perform the Fourier Transform of $U({\bf q})$  
\begin{equation}
U({\bf x})=\int \frac{d^3 q}{(2\pi)^3}  U({\bf q})e^{i{\bf q\cdot \bf x}}.
\end{equation}

The result can be written in the form of Eq. (\ref{1}) as follows: 

\begin{equation}
U({\bf{p_1}},{\bf{p_2}},{\bf{x}}) = V_c(r) + V_1(r) O_1 + V_2(r) O_2 + V_3(r) O_3 , \label{massive_vector_boson_potential}
\end{equation}
where:
\begin{eqnarray}
 V_c(r)  =  \left[ 1 + \frac{Q_c^2}{8c^2}\left(\frac{1}{m_1^2} + \frac{1}{m_2^2} \right) \right]\frac{e^{-Q_c r}}{4 \pi r}- \frac{1}{8c^2} \left(\frac{1}{m_1^2} + \frac{1}{m_2^2} \right) \delta ( {\bf{x}})
\end{eqnarray}

\begin{eqnarray}
 V_1(r) O_1 = \frac{1}{4m_1 m_2 c^2} \left[  \delta(\mathbf{x}) \left(-1 +\frac{4 \pi}{3} \right) + \frac{e^{-Q_c r}}{4 \pi r} \left( Q_c^2 + \frac{Q_c}{r} + \frac{1}{r^2} \right) \right]  {\bf{\sigmavec}_1} \cdot \mathbf{\sigmavec_2}
\end{eqnarray}  

\begin{eqnarray}
 V_2(r) O_2 = - \frac{1}{4 \pi} \frac{1}{4 m_1 m_2 c^2} \frac{e^{-Q_c r}}{r^3} \left( Q_c^2 + \frac{3 Q_c}{r} + \frac{3}{r^2} \right) (\mathbf{\sigmavec_1} \cdot \mathbf{x})(\mathbf{\sigmavec_2} \cdot \mathbf{x})
\end{eqnarray}
and:
\begin{eqnarray}
 V_3(r) O_3 = \frac{e^{-Q_c r}}{4 \pi r^2} \left( Q_c + \frac{1}{r} \right) \left[ - \frac{\mathbf{\sigmavec_1} \cdot(\mathbf{x} \times \mathbf{p_1})}{(2m_1 c)^2}  + \frac{\mathbf{\sigmavec_2} \cdot(\mathbf{x} \times \mathbf{p_2})}{(2m_2 c)^2}  + \frac{\mathbf{\sigmavec_1} \cdot(\mathbf{x} \times \mathbf{p_2})}{2 m_1 m_2 c^2}  - \frac{\mathbf{\sigmavec_2} \cdot(\mathbf{x} \times \mathbf{p_1})}{2 m_1 m_2 c^2}  \right].
\end{eqnarray}

For the scalar exchange boson interacting by $ \bar{\psi} \psi \phi$, the two-body potential is straightforward an similar to above computation:

\begin{eqnarray}
 \lefteqn{U({\bf p_{1}},{\bf p_{2}},{\bf q})= } \nonumber\\
   &=&    \Biggl[ \frac{1}{{\bf q}^2+Q_{c}^2} -\frac{1}{8c^2}\left(  \frac{1}{m^2_{1}}+ \frac{1}{m^2_{2}} \right)  \frac{{\bf q}^2}{{\bf q}^2+Q_{c}^2} - \frac{1}{2 m_1^2 c^2} \frac{\mathbf{p_1 \cdot ( p_1 + q)}}{\mathbf{q^2} + Q_c^2} - \frac{1}{2 m_2^2 c^2} \frac{\mathbf{p_2 \cdot ( p_2 - q)}}{\mathbf{q^2} + Q_c^2}  \nonumber \\  
   && {}-\frac{i}{(2m_{1}c)^2}\frac{{\bf \sigmavec_{1}} \cdot [{\bf q} \times {\bf p_{1}}]}{{\bf q}^2+Q_{c}^2} + \frac{i}{(2m_{2}c)^2} \frac{{\bf \sigmavec_{2}} \cdot [{\bf q} \times {\bf p_{2}}]}{{\bf q}^2+Q_{c}^2} \Biggr]. \nonumber \\
\end{eqnarray}

We note that in this case again the lowest order is the Yukawa interaction, nevertheless, there are neither spin-spin nor tensor interactions. 

   \section{Examples of Sommerfeld Enhancement}     

The Sommerfeld enhancement is a quantum mechanical effect, which leads to an increase of the total scattering cross section. In order to compute this effect 
we need to solve the radial Schr\"odinger equation, which for a particle of mass $\mu$ in presence of a central potential $V(r)$, is given by
\begin{equation}
-\frac{d^2 \chi(r)}{dr^2}+\left(  2\mu (V(r)-E) -\frac{l(l+1)}{r^2}   \right) \chi(r)=0. \label{Sch_radial_eq}
\end{equation}
where $R(r)=\frac{\chi(r)}{r}$ is the radial part of the wave function. The wave function satisfies the boundary conditions $\chi(0)=0$ and $\chi(r) \to \sin(kr +\delta)$ as $r \to \infty$.

The Sommerfeld enhancement of the scattering cross section then is computed by \cite{arkani} 
\begin{equation}
S_k=\left| \frac{\frac{d\chi(0)}{dr}}{k}   \right|^2 \label{Sommerfeld_factor}
\end{equation}
or equivalently  by 
\begin{equation}
S_k=\left| \frac{ R(\infty)}{R(0)}  \right|^2.
\end{equation}

Below we study some examples of Sommerfeld enhancement using nuclear potentials:

\subsection{Yukawa Potential for Fermions}

In the two-body Breit equation for massive vector boson computed in the above section, we consider only large terms contributions of the potential (\ref{massive_vector_boson_potential}). Thus, the first relevant term for the Sommerfeld enhancement is the spin-spin interaction, {\it v.i.z} 

\begin{equation}
U(\mathbf{x}) = (a + b \mathbf{ \sigmavec_1 \cdot \sigmavec_2})\frac{e^{- Q_c r}}{r}.
\end{equation}

Therefore, we have a Yukawa bosonic  potential ($b = 0$) and in this case the Schr\"odinger equation is solved by using a tensor product ansatz  between the position and spin space. 

If we work in the momentum-representation, the only non-diagonal part of the hamiltonian is the spin-spin interaction $\mathbf{ \sigmavec_1 \cdot \sigmavec_2}$. which can be diagonalized in the common base $\left| s_1 m_1 \right> \otimes \left| s_2 m_2 \right>$. 

After this, we obtain two equations, one for the triplet states $\psi_t$ and one for the singlet state $\psi_s$:

\begin{equation}
 \left[ \frac{\mathbf{p}^2}{2 \mu} +  (a + b)\frac{e^{- Q_c r}}{r} \right] \psi_t = E \psi_t
\end{equation}

\begin{equation}
 \left[ \frac{\mathbf{p}^2}{2 \mu} +  (a - 3 b)\frac{e^{- Q_c r}}{r} \right] \psi_s = E \psi_s
\end{equation}

Therefore in this case, we have to solve the well-know problem of a particle in a Yukawa potential of the form $\gamma e^{- Q_c r}/r$ with $\gamma = a + b $ and $\gamma = a - 3 b$ , for the triplet and singlet state, respectively. It is well know that the total cross section for this nucleon scattering is given by $\sigma=\frac{1}{4}\sigma_s+\frac{3}{4}\sigma_t$. Where $\sigma_s$ and $\sigma_t$ are the cross section for the singlet and triplet state, respectively.
Hence we can note that Sommerfeld enhancement factor is given by:

\begin{equation}
S = \frac{1}{4}S_s + \frac{3}{4} S_t
\end{equation} 

Where $S_t$ and $S_s$ are the enhancement factor of the particle in the Yukawa potential with $\gamma = a + b $ and $\gamma = a - 3 b$, respectively.
Finally we can note that in the massive scalar boson case we do not have this spin-spin term ( $\gamma = a$)

\subsection{Exponential Potential}

As was discussed in section II, the form of the nucleon-nucleon potential is not know in exact form, sometimes we have choose well phenomenological potentials in order to describe such interaction. In this sense, $$V(r)= V_0 e^{-Q_cr} $$ can be used as a nucleon potential approximation \cite{Davydov},  thus, this could another possible way for dark matter interaction and for this reason we investigate below whether this potential exhibits Sommerfeld enhancement.

In order words  we need to compute the wave function which is solution of the Schr\"odinger radial equation (\ref{Sch_radial_eq}) and since we are interested in low energy processes we need  only s-waves ({\it i.e.} solutions with $l=0$).

In order to compute $\chi(r)$ we perform the standard change of variable
\begin{equation}
x=\alpha e^{-\beta r},
\end{equation}
where $\alpha$ and $\beta$ are constants, and choosing this constants as 
\begin{equation}
\alpha=\frac{2}{Q_c}\sqrt{2\mu V_{0}},  \hspace{5 mm} \beta=\frac{Q_c}{2}.
\end{equation}
we can rewrite Eq.(\ref{Sch_radial_eq}) in the form of a Bessel equation:
\begin{equation}
\frac{d^2 \chi(x)}{dx^2}+ \frac{1}{x}\frac{d \chi(x)}{dx}+  \left(1-\frac{\rho^2}{x^2}\right)  \chi(x)=0,
\end{equation}
where $\rho=2i|k|/Q_c$, corresponds to states with positive energy $E$.

The general solution is given by
\begin{equation}
\chi(x)=A J_{\rho}(x)+ B N_{\rho}(x),
\end{equation}
with $ J_{\rho}$ and $N_{\rho}$ are the Bessel and Newman function respectively. With the requirement  that the wave function is finite as $r \to \infty$ we conclude $B=0$.

The Sommerfeld enhancement Eq (\ref{Sommerfeld_factor}) becomes 
\begin{equation}
S_{k}=\left| -\frac{\alpha \beta}{k}  \frac{d J_{\rho} (x=\alpha)}{dx}  \right|^2 . \label{som}
\end{equation}

%

Using the asymptotic form of the Bessel function  \cite{Abramowitz}, $z \gg 1$ and $|z| \gg |\rho^2-\frac{1}{4}|$, {\it i.e.} in the low energy limit (or equivalently $V_{0} \gg \frac{k^2}{2\mu}$), we have

\begin{equation}
J_{\rho}(z)=\sqrt{{\frac{2}{\pi z}}}e^{\nu \pi/2} \cos \left(z-\frac{\pi}{4}\right),
\end{equation}
where $\rho=i\nu$.

Using this last result and (\ref{som}) the Sommerfeld enhancement factor becomes 

\begin{equation}
S_{k}=\frac{2 \beta^2 \alpha}{\pi k^2} e^{ \pi  k/ \beta}\sin^2 \left(   \alpha-\frac{\pi}{4}              \right).
\end{equation}

   \section{Discussions and Conclusions} 
In this paper we have proposed a  point of view inspired in nuclear physics in order to study the fermionic dark matter dynamics. This perspective is particularly interesting because, in principle, would allows a dictionary between   \lq \lq nucleon $\leftrightarrow$ dark matter" clear and directly. 

Specifically, we obtained the following results. 1) we have computed, by using the $1/c$, the two-body potential for scalar and vector interactions and we have derived the corresponding Yukawa potential; 2) If the process involve fermion and vector, the two-body potential has the form 
$(a + b \mathbf{ \sigmavec_1 \cdot \sigmavec_2})e^{- Q_c r}/r$ and therefore, the modification of the Sommerfeld enhancement is given by $S = \frac{1}{4}S_s + \frac{3}{4} S_t$ due to the additional spin-spin term $\mathbf{ \sigmavec_1 \cdot \sigmavec_2}e^{- Q_c r}/r$ in the potential $U(\mathbf{x})$; 3) 
On the other hand, if the process is through a scalar boson the relevant long range term is pure Yukawa term $\alpha e^{- Q_c r}/r$, so we do not have the above modification to the enhancement factor.

Finally, we have also considered an attractive  phenomenological nuclear potential $- V_o e^{-Q_c r}$ and show explicitily as emerges  a Sommerfeld enhancement factor, so in principle, this potential also could used to model processes of enhancement and dark matter. 

\begin{acknowledgments}
This work was supported by grants from CONICYT 21140036 (D.C) (J.G), CONICYT 21090138 (A.R) and FONDECYT-Chile 1130020 (J. G.). 
\end{acknowledgments}

\end{document}